\documentclass{acmart}
\AtBeginDocument{%
  \providecommand\BibTeX{{%
    \normalfont B\kern-0.5em{\scshape i\kern-0.25em b}\kern-0.8em\TeX}}}

\setcopyright{acmcopyright}
\copyrightyear{2024}
\acmYear{2024}
\acmDOI{XXXXXXX.XXXXXXX}

\acmConference[CCAI 2024]{CHI 2024 Workshop on Child-centred AI Design}{May 11,
  2024}{Honolulu, HI, USA}
\acmBooktitle{CHI 2024 Workshop on Child-centred AI Design, May 11, 2024, Honolulu, HI, USA} 
\begin{document}

\title{Reimagining AI: Exploring Speculative Design Workshops for Supporting BIPOC Youth Critical AI Literacies}


\author{Sadhbh Kenny}
\affiliation{%
  \institution{Simon Fraser University}
  \city{Surrey}
  \state{BC}
  \country{Canada}
}
\email{sadhbh_kenny@sfu.ca}
\orcid{XXXX-XXXX-XXXX}

\author{Alissa N. Antle}
\affiliation{%
  \institution{Simon Fraser University}
  \city{Surrey}
  \state{BC}
  \country{Canada}
}
\email{aantle@sfu.ca}
\orcid{0000-0001-9427-8772}


\begin{teaserfigure}
 \includegraphics[width=\textwidth]{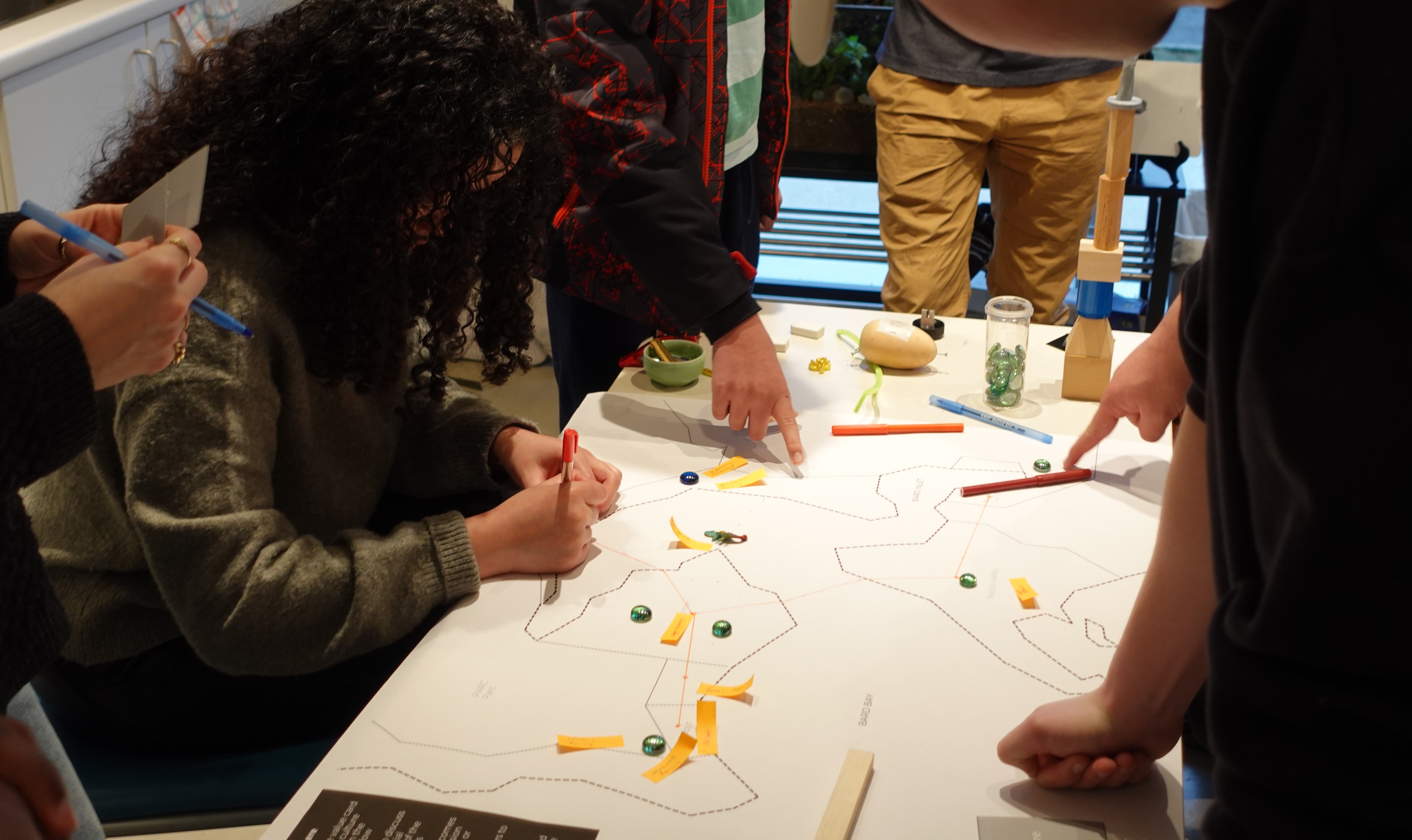}
  \caption{World-building activity .}
  \Description{Children gathered around map, pointing to various places and discussing possible reconfiguration.}
  \label{fig:teaser}
\end{teaserfigure}

\begin{abstract}\textbf{ABSTRACT}

\textbf{Background:} As Artificial Intelligence (AI) ecosystems become increasingly entangled within our everyday lives, designing systems that are ethical, inclusive, and socially just is more vital than ever. It is well known that AI can have algorithmic biases that reflect, extend, and exacerbate our existing systemic injustices \cite{benjamin_race_2023,winner_artifacts_1980,noble_algorithms_2018}. Yet, despite most teenagers interacting with AI daily \cite{ito_algorithmic_2023}, only few have the opportunity to learn how it works and more importantly its socio-technical complexities. This is a particularly salient issue for communities that have been historically marginalized in technology discourse. Not only are BIPOC teens misrepresented throughout AI development and implementation \cite{buolamwini_gender_2018}, but they are also significantly less likely to receive STEAM education \cite{wang_is_2017}.
 
\textbf{Method:} In response to these unprecedented socio-technical challenges and calls for more critical approaches to child-centered AI design and education \cite{iivari_critical_2022, dindler_computational_2020,ko_it_2020} , we explore how we can leverage co-speculative design practices \cite{bleecker_manual_2022,auger_speculative_2013,dunne_speculative_2013} to help scaffold BIPOC youth critiques of existing AI systems and support the critical re-imagining of more just AI futures. Drawing on Haraway’s concepts of \textit{Situated Knowledges} \cite{haraway_situated_1988} and \textit{Speculative Fabulations} \cite{haraway_speculative_2011}, these workshops aim to highlight the unique ways in which historically marginalized youth perceive AI as having social and ethical implications and how they envision alternative worlds with AI. The ability to challenge dominant discourse and envision how future AI implementations may impact society are key competencies of critical digital literacy \cite{freire_pedagogy_2012,van_sluys_researching_2006}; however, the use of tangible world building approaches to speculative design and critical AI literacies have been relatively under -explored in CCI. 

%What are key elements of research methods (vs workshop design)? What is RQ?

Conventionally, mapping involves the delineating boundaries, control, and ownership. This process often conveys a symbolic representations of relationships among objects, actors and space; consequently, serving as a tool for constructing knowledge and shaping reality\cite{geerts_navigating_2022}. Historically, cartography has been used for political domination and colonial endeavors. We appropriate and subvert this practice as means of ontological and epistemological world-building \cite{geerts_navigating_2022} and a tool for counter-storying alternative AI realities. This ongoing work uses the openness of speculative cartography as a catalyst for reflection, disorientation and the reimagination of how we could co-exist with “intelligent” systems in a world absent of techno-capitalist values. 

Our case study describes three 2 hour sessions of a larger 8 week black-led AI STEAM program. In-person sessions include a combination of  hands-on, speculative and participatory learning activities that focus on AI ethics.  Drawing on pre-post surveys, workshop recordings and field notes, we provide initial insights on the following research questions: RQ1:How do youth (grades 9 -12) belonging to minority backgrounds perceive the social and ethical implications of their everyday AI technologies? RQ2: In what ways can co-speculative design workshops help youth to cultivate critical AI literacies? RQ3: In what ways can co-speculative design workshops support the reimagination of alternative AI realities?

\textbf{Contribution:} 
% What is the knowledge contribution? (and social contribution)
% new methods for helping youth understand the complexities of AI ethical implicaitions 
% New Worlds, New problems, New Artifacts
% help youth understand the cycical nature of the technoculture and encouraging the critical speculation of how things could be different. 
% highlight the important of care...? thinking about different viewpoint (aside from their own) 

% identity + implications for others ( at different social ecological levels) 

%using value sensitive design - but going beyond these reductive questions and bringing in more complex questions 

The contributions of this work will be 1) a discussion of how youth perceive AI as having social and ethical implications 2) an exploration and nuanced understanding of how speculative approaches can be leveraged to support children's engagement with complex socio-technological issues and 3) enable youth to take ownership of and critically reflect on the socio-political implications of their own AI designs. Overall, we hope these workshops can empower youth to take on new bias towards action, resistance and collective dreaming.

\end{abstract}


\maketitle
%%WORKSHOP INFO
%AI systems and algorithms have become central to children’s digital landscapes, seamlessly integrating into connected toys, smart home IoT devices, and daily-use apps. Their prevalence is attributed to their ability to enhance user experiences, making them adaptive, compelling, and personalized. During our first CCAI workshop at CHI 2023, we focused on conceptualizing child-centered AI and highlighted its significance. This year, our second CCAI workshop will pivot to the hands-on application of these conceptual foundations in child-centered AI. We will investigate how next-gen technologies like generative AI/LLMs and personal robots can be designed from a child-centric perspective. We invite papers, up to 4 pages excluding references, covering: predominant challenges, methodologies employed, foundational principles, practical safeguards, and insights for translating child-centered AI concepts into practice. Submissions should conform to the CHI Extended Abstract format and be submitted to our email at \textbf{ccai2024chi@gmail.com}. The submissions will be reviewed by the workshop organizers and program committee. All accepted papers will be published through our website and presented at the workshop, provided at least one author attends the workshop and registers at least one day of the conference.

\section{Why do you want to participate?}
We are researchers from the Tangible Embodied Child-Computer Interaction (TECI) Lab at Simon Fraser University located in Vancouver, Canada: Alissa N. Antle (Lab Director and Professor) and Sadhbh Kenny (M.A. Student). Our lab conducts design-oriented research to design, build and evaluate technology that improves, augments, and supports children’s cognitive and emotional development. More recently, we have been developing novel ways of engaging children in participatory and co-design methods to include them as design partners in the development of their emerging technologies, such as AI \cite{kitson_co-designing_2023, antle_there_2022}.  

We are interested in discussing potential uses of participatory and speculative design methods to:  1) investigate ethical issues surrounding AI technologies for and with marginalized youth; 2) inform our research, design and development of AI to ensure it is situated in and inclusive of youths' lived experiences; and 3) explore ways to cultivate youth critical technical literacies for emerging technologies, such as AI. We would also like to share our relevant past and current research on existing child-centered AI design methods. 

\section{What can you contribute?}
First, we would be happy to share  our current research that explores using co-speculative design methods with BIPOC youth to engage them in critical digital literacies and re-imagining more preferable AI futures (see above abstract for brief description). In this work we are investigating some of the ways that we can leverage speculative design to situate methods in BIPOC lived experiences, elicit youth socio-political critiques of AI systems, and provide scaffolding to disrupt our normative ways of using and envisioning AI \cite{bleecker_manual_2022,auger_speculative_2013,dunne_speculative_2013}. 

Second, we would be happy to share our findings from a recent rapid review that synthesized the existing literature on child-centered design methods focusing on the ways that children and youth have been involved in AI research and development. Following Druins \cite{druin_cooperative_1999} child centered design roles and Barengret et al. \cite{barendregt_role_2016} design stages, we distilled and extracted the ways in which various methods enabled children’s involvement in the sensitizing, requirements, design, and evaluation of AI systems. Findings include highlights of  best practices for involving children in the design and implementation of AI systems.

\section{What do you want to get out of the workshop?}
In participating in this workshop, we hope to share and receive feedback on our current research with youth, AI design and review paper. We would like to learn from others about best practices in involving children and youth in the design and implementation of AI systems, and discuss the potential of co-speculative design to explore the ethical issues related to emerging technologies. We also welcome potential research collaborations around this topic.

\bibliographystyle{ACM-Reference-Format}
\bibliography{chioverleafref2}

\end{document}